# ADAPTIVE DATA STREAM MANAGEMENT SYSTEM USING LEARNING AUTOMATA

Shirin Mohammadi[1], Ali A. Safaei[2], Fatemeh Abdi[3] and Mostafa S. Haghjoo[4]

[1]Department of Computer Engineering, Iran University of Science and Technology, Tehran, Iran
*sh_mohammadi@comp.iust.ac.ir*

[2]Department of Computer Engineering, Iran University of Science and Technology, Tehran, Iran
*safaeei@iust.ac.ir*

[3]Department of Science, Babol-Branch, Islamic Azad University, Babol, Iran
*sulmaz_abdi@yahoo.com*

[4]Department of Computer Engineering, Iran University of Science and Technology, Tehran, Iran
*haghjoom@iust.ac.ir*

## ABSTRACT

*In many modern applications, data are received as infinite, rapid, unpredictable and time- variant data elements that are known as data streams. Systems which are able to process data streams with such properties are called Data Stream Management Systems (DSMS). Due to the unpredictable and time-variant properties of data streams as well as system, adaptivity of the DSMS is a major requirement for each DSMS. Accordingly, determining parameters which are effective on the most important performance metric of a DSMS (i.e., response time) and analysing them will affect on designing an adaptive DSMS.*

*In this paper, effective parameters on response time of DSMS are studied and analysed and a solution is proposed for DSMSs' adaptivity. The proposed adaptive DSMS architecture includes a learning unit that frequently evaluates system to adjust the optimal value for each of tuneable effective. Learning Automata is used as the learning mechanism of the learning unit to adjust the value of tuneable effective parameters. So, when system faces some changes, the learning unit increases performance by tuning each of tuneable effective parameters to its optimum value. Evaluation results illustrate that after a while, parameters reach their optimum value and then DSMS's adaptivity will be improved considerably.*

## 1. INTRODUCTION

Database management systems (DBMSs) simply store huge amounts of data. When users of applications need to use huge amounts of data, they once load them from a DBMS or retrieve a specific part of them by a simple real time query. The DBMSs are the best in executing ad-hoc queries.

But various applications need to process data streams. The data streams include infinite, continuous, rapid and timely variant data elements [1, 2]. These applications need a new class of systems called as data stream management systems (DSMSs). The DSMSs are able to propose queries on data streams. These queries are executing in a long time mode since data arrival in a continuous mode, and are called as continuous queries [1].

There are bold challenges about DSMSs. Dynamic features of system, specifications of input data streams, the output and queries while executing the continuous queries, all may change significantly. When each of the above categories changes a single execution which has the best performance before the change may suddenly change to the worst performance one. Thus, adaptivity for DSMSs is an important issue, because regardless of it the performance may decrease significantly while time is going on.

Some performance metrics of DSMSs which have to be evaluated regularly are as [3, 4]:

1- Response time (tuple latency): the amount of time (or average time) which a tuple (or an order of them) has to pass to execute a query. Of course this time includes all waiting times in buffers.

2- Memory usage: the maximum amount of memory which is used by the system.

3- Throughput: number of output tuple in a specific unit of time.

4- Accuracy of results: rate of accuracy or correctness of results in fault tolerant conditions.

5- Tuple loss: ratio of lost tuples that total count of input tuples.

These metrics are not independent of each other. For example reducing the tuple latency usually increases the amount of memory usage or reducing the memory usage leads to loss of tuples and finally reduction in accuracy of results. Selecting a metric can significantly influence the other ones. So tradeoff has to establish between performance metrics. A data stream system establishes a rational tradeoff based on application needs [3].

The rest of the paper is organized as follow: effective parameters on DSMSs' response time are reviewed and analysed in section 2. The proposed adaptive DSMS is presented in section 3; effective parameters are considered to be tuneable or not, and learning automata used for adjusting tuneable effective parameters in the proposed adaptive DSMS is explained in detail. Performance evaluation is discussed in section 4 and related work is reviewed in section 5. Finally, we conclude and propose some future work in section 6.

## 2. BACKGROUND

The most important performance metric for DSMSs is the response time. Response time of a query in a data stream system depends on several parameters. Some are less effective whilst some are more.

According to the figure 1, effective parameters on response time of a DSMS are represented in total categories of: 1) *Data stream properties*, 2) *Query properties*, 3) *Query execution properties*, 4) *Output properties*, 5) *System properties (static conditions)* and 6) *System condition (Dynamic)*.

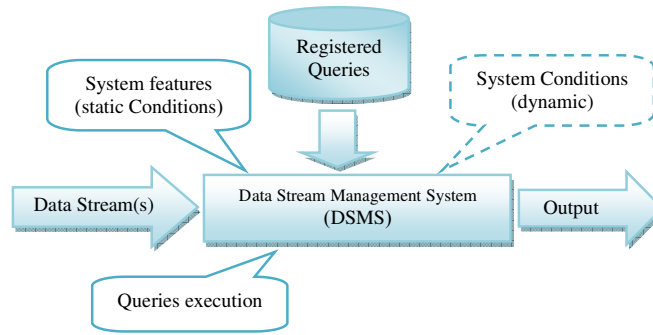

Figure 1. Categorizing the effective parameters on response time [23]

- **Data Stream Properties**

A data stream includes data elements generated in an infinite, continuous and rapid manner which varies in time. In other words data stream of $S$ is a set of $s$ elements with time stamp of $\tau$ which the elements arrive to the system in time stamp order. The time stamp specifies the logical entrance time of a tuple into the data stream.

Data streams are generated by external resources or other applications and are sent to the DSMS. DSMSs have no direct access or any kind of control on data resources [3]. The most important features of a data stream which are effective on response time of the system include: *1) Type of elements in a data stream*, *2) Domain of attributes*, *3) Number of attributes*, *4) Data stream distribution* and *5) Arrival rate into the system*.

- **Query properties**

Continuous queries are those which have several processes on new data to generate new results. They are executed in a long lasting mode and generate the results continuously [3].
In this category details of effective parameters on response time are considered as: *1) Type of query*, *2) Number of operators in a query*, *3) Arrangement of operators in query plan* and *4) Type of operators*.

- **Execution of Queries**

To execute multiple queries concurrently in a DSMS, first the query selection has to be executed. In second step the manner of accurate execution of queries has to be specified. The first and second steps are discussed as scheduling queries and scheduling query operators [5]. Execution algorithm of operators is also an important issue about the response time of the system. Effective parameters of this category include: *1) Scheduling queries*, *2) Scheduling query operators* [6, 7] and *3) Scheduling algorithms of operators* [10].

- **Output properties**

This category includes parameters such as *1) Type of output* and *2) Number of outputs*.

- **System properties (static conditions)**

Depending on total or the static conditions of the DSMS, effective parameters on response time in this category can be considered as: *1) Number of registered queries*, *2) Query registration time* [8], *3)Number of processes (logical machines)* [9], *4) Processing capacity of processor(s)*, *5) Processor's architecture* [10] and *6) The amount of available memory*.

- **System condition (Dynamic)**

Dynamic status of the system includes changes which may occur while systems execution. This category includes parameters such as: *1) Allocated processing capacity of processor(s) for query execution*, *2) Memory usage*, *3) Overload in DSMS* [8, 11] and *4) Occurring deadlock*.

## 3. ADAPTIVE DSMS USING LEARNING AUTOMATA

At first, effective parameters which are not tuneable must be determined and removed from are discussion since we are not able to change them.

### 3.1. Effective Parameters on Response Time of a DSMS

Effective parameters on response time can be categorized as: 1) constant value parameters, 2) effective and non-tuneable parameters and 3) effective and tuneable parameters.

- **Constant Value Parameters**

This category includes parameters which are assumed with constant values of Table 1.

Table 1.  Parameters assumed with constant values

| Parameter Name | Symbol | Parameter's Value |
|---|---|---|
| *Type of data stream's elements* | $T_e$ | Tuple[*] |
| *Domain of attributes* | $D_A$ | Long[*] |
| *Number of attributes* | $N_A$ | Five[*] |
| *Type of query* | $T_Q$ | Continues query |
| *Scheduling Operators* | $S_{Op}$ | FIFO[**] |
| *Algorithm of operators* | $A_{Op}$ | Developed algorithm in the system |
| *Type of output* | $T_{Out}$ | Relation |
| *Number of outputs* | $N_{Out}$ | One |
| *Query registration time* | $R_Q$ | Pre-Defined |

- **Effective and non-tuneable Parameters**

Parameters of table 2 are effective on response time of the DSMS but they have non-tuneable values. The reason of being non-tuneable for each one is provided in Table 2.

Table 2.  Effective and non-tuneable parameters

| Parameter Name | Symbol | Reason |
|---|---|---|
| *Data stream distribution* | $F$ | The DSMS has no control on arrival order, arrival rate or on distribution of the input stream. Because the data streams of distributed resources are reaching the system remotely [2]. |
| *Arrival rate into the system* | $\alpha$ | |
| *Processing capacity of processor(s)* | $C_p$ | In cause of limitations in resources such as processor(s) and memory, these parameters are assumed as non-tuneable ones. |
| *System architecture* | $A_p$ | |
| *The amount of available memory* | $M_A$ | |
| *Allocated processing capacity of Processor(s) for query execution* | $C_{Dp}$ | Considering that system processor(s) may simultaneously be owned to other processes while executing queries, then always just a part of processor(s) is assigned to the process of query which of course this parameter is an non-tuneable one for the system. |

---

* The Input data set includes data of monitoring IP packets which is located in Internet Traffic Archive (ITA) [12]. One of traces, specifically the "DEC-PKT" contains all wide-area traffic of an hour between Digital Equipment Corporation and the rest of the world. This real-world data set is used in our experiments. Two types of monitored packets, the TCP and the UDP packets, are selected as input streams. Each TCP packet contains five items of source address, destination address, source port, destination port, and length. UDP packets are the same as TCP missing the length of packets.

** considering the results of experiments on scheduling algorithms [6], regarding the simplicity of implementation and less overhead for the system, the FIFO is the best choice.

| | | |
|---|---|---|
| *Overload in dSMS* | *Ov* | Large number of resources is one of the features of the DSMSs which the rate of data arrival in them can unpredictably be high. While premature data arrival, available demands on system resources (like the processor, main memory or the band width) may exceed of the capacity. That condition is called as occurring overload in the system [8, 11]. This parameter is also an non-tuneable one. |
| *Occurring deadlock* | *D* | If three sufficient conditions for occurring deadlock are established through the data stream system, the failure occurs and the response time will be infinite. occurring deadlock is also an non-tuneable parameter. |

- **Effective and Tuneable Parameters**

Table 3 represents effective and tuneable parameters. Three parameters of number of operators in a query, arrangement of operators in query plan and type of operators are studied in topics such as optimizing query plans and adaptivity of query processing [2]. About four parameters of scheduling queries, number of registered queries, number of processes and memory usage we can use the learning unit for optimal setting of values along the execution of the system, dynamically.

Table 3.  Effective and tuneable parameters

| Parameter Name | Symbol | Reason |
|---|---|---|
| *Number of Operators in a Query* | $N_{Op}$ | Count, type and order of operators in a query can set to be optimum. These three parameters are fully studied in topics such as query plan optimization and adaptivity of query process. |
| *Arrangement of Operators in Query Plan* | $QP$ | |
| *Type of Operators* | $T_{Op}$ | |
| *Scheduling Queries* | $S_Q$ | The optimum scheduling algorithms for DSMSs are ones which work cyclic like the Round Robin. In these algorithms the most important parameter is the value of Time Quantum, which can simply set to the optimum value regarding feedbacks from the environment. |
| *Number of Registered Queries* | $N_Q$ | Considering the priority of registered queries by the user, while reducing performance factors, queries with less priority can temporarily suspend and reload in proper time. |
| *Number of Processes (logical Machines)* | $N_p$ | In relation to the processing capacity of processor(s), number of in process |

| | | queries and number of processes of the system can be adjusted dynamically. |
|---|---|---|
| *Memory Usage* | $M_U$ | By increasing the used memory, tuples have to wait more in queues. Consequently the response time of the system will increase. On the other hand decreasing the used memory forces the system to dispose tuple which finally leads to decrease the Accuracy of results. Dynamic setting of used memory can establish a proper balance between response time factor and loss of tuples rate. |

### 3.2. Improving DSMS Adaptivity by using Learning Automata

Learning can be used as a solution to establish adaptivity in systems. By using the learning in a proper position of the system, each constructor parts of it, even in receiving incomplete and non-deterministic information, can gradually achieve the required optimum strategy of control based on defined factors about it.

Learning automata as an abstracted model, is able to do some limited operations. Each selected activity is evaluated by a probable environment and respond the learning automata. The learning automata use this response and select its activity for next step [24, 25]. Figure 2 represents the relation between the learning automata and the environment.

The environment can be shown by triple set of $E = \{\alpha, \beta, c\}$ which set of inputs as $\alpha = \{\alpha_1, \alpha_2, \cdots, \alpha_r\}$, set of outputs as $\beta = \{\beta_1, ..., \beta_m\}$ and set of fine possibilities are shown as $c = \{c_1, c_2, ..., c_r\}$. When $\beta$ is a set with two members then the environment is in type of P. In such environment $\beta_1 = 1$ is considered as a fine and $\beta_2 = 0$ is considered as reward. In an environment of Q type, $\beta(n)$ can be discretely as a value between [0, 1] and in an environment of type S, then $\beta(n)$ is a random variable in range of [0, 1].

The $c_i$ is probability of having unfavourable results for $\alpha_i$ action. In a static environment $c_i$ values remain unchanged, while in a non-static environment these values change along the time. A learning automaton is separated into two groups with constant and variable structures [24, 25].

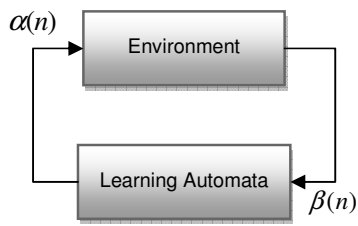

Figure 2. Relation between learning automata and environment

Learning automata with variable structure is represented by quadruple set of $\{\alpha, \beta, p, T\}$ which $\alpha = \{\alpha_1, ..., \alpha_r\}$ is set of actions for automata, $\beta = \{\beta_1, ..., \beta_m\}$ is set of inputs for automata and $p = \{p_1, ..., p_r\}$ is the vector of probability of selection of each of actions and $p_{n+1} = T[\alpha_n, \beta_n, p_n]$ is the learning algorithm. In these kinds of automata, if the $\alpha_i$ in nth stage of selection receives desirable (or undesirable) response from the environment, probability of $p_i(n)$ increases (or decreases) and other probabilities decrease (or increase). The changes occur so that sum of $p_i(n)$ s is always constant and equals to one [24, 25].

### 3.3. Architecture of the Proposed DSMS

In Figure 3 as the parallel query processor engine, we have *k* processors (logical machines) which are in relation to the others. These machines can be physical machines (like processors of a multi-processing system or nodes of a clustered computer) or virtual machines (like threads which are executed on cores of multi core processor).

For parallel execution of query on *k* logical machines, for each registered query on the represented system, first query plan has to be established. Then k same copies of this query plan are generated and each of them are sent to one of k logical machines. The goal of this is to inform all machines about the query plan (operators and their order), but the issue that each machine has to execute which operator and how to communicate with other machines so that the query plan executed on all machines in parallel mode, is specified by the scheduling algorithm.

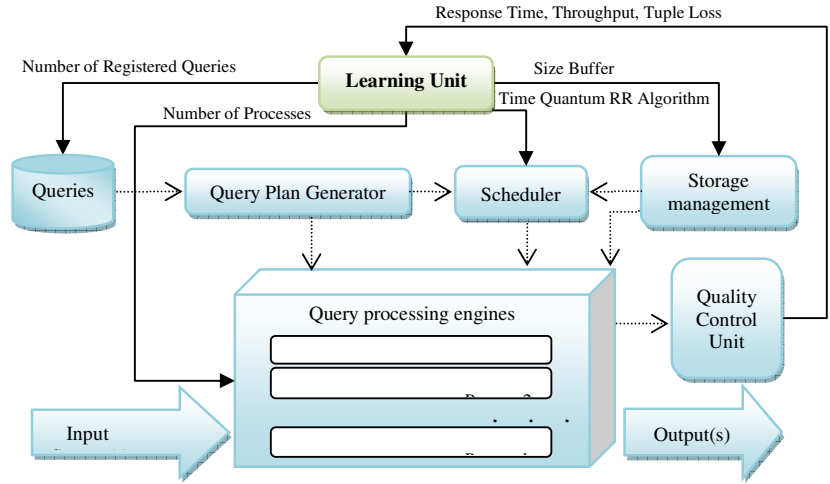

Figure 3. Architecture of the adaptive DSMS

For dynamically setting of the effective and tuneable parameters as mentioned below, the learning unit is used in this system.

- Time quantum RR algorithm ( $Q_{RR}$ )
- Number of registered queries ( $N_Q$ )
- Number of processes ( $N_P$ )
- Size buffer ( $S_B$ )

As we can see in figure 4, the learning unit returns three values of below by getting feedback from the quality control unit and consequently regulate four mentioned parameters dynamically.

- Response time ( $R_t$ )
- Throughput ( *Th* )
- Tuple loss ( $T_{loss}$ )

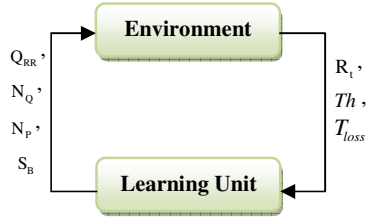

Figure 4. Learning Unit

Learning unit is a leaning automaton with variable structure which is defined by quadruple set of $\{\alpha, \beta, p, T\}$. $\alpha$ is set of automata actions as a set of quadruple elements of increase ($\uparrow$), decrease ($\downarrow$), unchanged (-) and parameters of time quantum of RR algorithm to schedule between Queries ($Q_{RR}$), number of registered queries ($N_Q$), number of processes ($N_P$) and size of buffer ($S_B$). The $K^{\uparrow}$ represents amount of increase, $K^{\downarrow}$ as amount of decrease and $f$ as a factor of addition (subtraction) or multiplication (division), represents increase (decrease) of the X parameter. For example if action of ($\downarrow Q_{RR}, N_P, N_Q, \uparrow S_B$) is selected, it means that $Q_{RR}$ parameter subtracts or divides from the $f_{Q_{RR}}$ in $K^{\downarrow}_{Q_{RR}}$ units. Values of $N_P$ and $N_Q$ parameters remain unchanged and $S_B$ parameter adds or multiplies to $f_{S_B}$ in $K^{\uparrow}_{S_B}$ units. Then $\alpha$ function is defined in form of below:

$$\alpha = \left\{ \begin{aligned} &(Act(Q_{RR}, K^{\uparrow}_{Q_{RR}}, K^{\downarrow}_{Q_{RR}}, f_{Q_{RR}}), Act(N_P, K^{\uparrow}_{N_P}, K^{\downarrow}_{N_P}, f_{N_P}), Act(N_Q, K^{\uparrow}_{N_Q}, K^{\downarrow}_{N_Q}, f_{N_Q}), Act(S_B, K^{\uparrow}_{S_B}, K^{\downarrow}_{S_B}, f_{S_B})) \\ &Act(X, K^{\uparrow}, K^{\downarrow}, f) \in \{\uparrow X, \downarrow X, X\} \end{aligned} \right\} \quad (1)$$

Considering four parameters, each one for each of the actions of addition, subtraction, multiplication and division, then $\alpha$ set has 81 members as:

$$\alpha = \left\{ \begin{aligned} &(\uparrow Q_{RR}, \uparrow N_P, \uparrow N_Q, \uparrow S_B), (\uparrow Q_{RR}, \uparrow N_P, \uparrow N_Q, \downarrow S_B), (\uparrow Q_{RR}, \uparrow N_P, \uparrow N_Q, S_B), \\ &(\uparrow Q_{RR}, \uparrow N_P, \downarrow N_Q, \uparrow S_B), (\uparrow Q_{RR}, \uparrow N_P, \downarrow N_Q, \downarrow S_B), (\uparrow Q_{RR}, \uparrow N_P, \downarrow N_Q, S_B), \\ &(\uparrow Q_{RR}, \downarrow N_P, \downarrow N_Q, \uparrow S_B), (\uparrow Q_{RR}, \downarrow N_P, \downarrow N_Q, \downarrow S_B), (\uparrow Q_{RR}, \downarrow N_P, \downarrow N_Q, S_B), \\ &(\downarrow Q_{RR}, \downarrow N_P, \downarrow N_Q, \uparrow S_B), (\downarrow Q_{RR}, \downarrow N_P, \downarrow N_Q, \downarrow S_B), (\downarrow Q_{RR}, \downarrow N_P, \downarrow N_Q, S_B), \\ &..., (Q_{RR}, N_P, N_Q, S_B) \end{aligned} \right\} \quad (2)$$

The $\beta$ is the input set of the learning unit which is defined as:

$$\beta = \{R, T_{loss}, Th\} \qquad (3)$$

The $p$ is probability vector of selecting each of the actions of the learning unit as:

$$p = \{p_1, p_2, ..., p_{81}\} \qquad (4)$$

If the $\alpha_i$ action is selected in nth stage and received desired answer from the environment, then the reward function of the learning algorithm is defined as:

$$\text{Reward} = \begin{cases} p_i(n+1) = p_i(n) + a[1 - p_i(n)] \\ p_j(n+1) = (1-a) p_j(n) \qquad \forall j \ j \neq i \end{cases} \qquad (5)$$

In this formula the $p_i(n)$ is probability of selection action of $\alpha_i$ in nth stage and $p_i(n+1)$ is probability of selecting action of $\alpha_i$ in **(n+1)**th stage. The **a** is the reward parameter and if after selecting the $\alpha_i$ action in nth stage, it receives an undesired response of the environment, the penalty function of the algorithm will be in form of below:

$$Penalty = \begin{cases} p_i(n+1) = (1-b)p_i(n) \\ p_j(n+1) = (\dfrac{b}{r-1}) + (1-b)\,p_j(n) \quad \forall j \ \ j \neq i \end{cases} \qquad (6)$$

The b is the penalty parameter and the r is count of set for $\alpha$ .

To specify desirability of the response from the environment, $f(n)$ function is defined as:

$$f(n) = \alpha_1 \times k_1 + \alpha_2 \times k_2 + \alpha_3 \times k_3 \qquad (7)$$

Which parameters of $K_1$, $K_2$, $K_3$ are defined in order of ratio of average response time than the response time of the nth stage, ratio of average tuple loss than tuple loss in nth stage and finally ratio of operational potency in nth stage than the average operational potency. $\alpha_1$, $\alpha_2$, $\alpha_3$ are coefficients between 0 to 1 which represent amount of importance for each of the parameters of $K_1$, $K_2$, $K_3$ which they are define as:

$$k_1 = \frac{\overline{R_i(n)}}{R_i(n)} \qquad (8)$$

$$k_2 = \frac{\overline{T_{loss}(n)}}{T_{loss}(n)} \qquad (9)$$

$$k_3 = \frac{Th(n)}{\overline{Th(n)}} \qquad (10)$$

Average response time $\overline{R_i(n)}$, average tuple loss $\overline{T_{loss}(n)}$ and average operational potency $\overline{Th(n)}$ in (n+1)th stage are defined as:

$$\overline{R_i(n+1)} = \gamma \times \overline{R_i(n)} + (1-\gamma) \times R_i(n) \quad , \quad \overline{R_i(0)} = 0 \qquad (11)$$

$$\overline{T_{loss}(n+1)} = \gamma \times \overline{T_{loss}(n)} + (1-\gamma) \times T_{loss}(n) \quad , \quad \overline{T_{loss}(0)} = 0 \qquad (12)$$

$$\overline{Th(n+1)} = \gamma \times \overline{Th(n)} + (1-\gamma) \times Th(n) \quad , \quad \overline{Th(0)} = 0 \qquad (13)$$

The F function for specifying desirability of the selected action in nth stage, works as below. If the value of $f(n)$ function is greater than or equal to $\overline{f(n)}$ value, the action is considered as desired action and if it is lower than $\overline{f(n)}$ the action is considered as an undesired one. So the function of learning algorithm of T calls the reward function in case of desired action and calls the *Penalty* function in case of undesired action.

$$F = \begin{cases} desirable & if \ \ f(n) \geq \overline{f(n)} \\ Undesirable & if \ \ f(n) < \overline{f(n)} \end{cases} \qquad (14)$$

The $\overline{f(n)}$ function is also defined as:

$$\overline{f(n+1)} = \gamma \times \overline{f(n)} + (1-\gamma) \times f(n) \quad , \quad \overline{f(0)} = 0 \qquad (15)$$

Along processing queries the learning unit starts to set values of these four parameters. After a while probability of action of $(Q_{RR}, N_P, N_Q, S_B)$, $p_{81}$ in learning unit reaches the constant value of one. This means that for next action there is no need to change values of parameters. Of course by changes in the DSMS, these values lose their optimum state and the learning unit after a while reached the optimum state in present condition by resetting the values. In this case resets the parameters adaptively and in present state.

## 4. PERFORMANCE EVALUATION

As the evaluation process we developed a prototype which been implemented in the Java language with JDK 6.0 on a machine which was equipped with a Core i7 2930 Intel processor and 6GB of RAM in Linux environment. The Input data set includes data of monitoring IP packets which is located in Internet Traffic Archive (ITA) [12]. One of traces, specifically the "DEC-PKT" contains all wide-area traffic of an hour between Digital Equipment Corporation and the rest of the world. This real-world data set is used in our experiments. Two types of monitored packets, the TCP and the UDP packets, are selected as input streams. Each TCP packet contains five items of source address, destination address, source port, destination port, and length. UDP packets are the same as TCP missing the length of packets.

Elements of the stream are in type of well-structures data (tuple). To schedule queries the Round Robin (RR) algorithm and to schedule query operators the FIFO algorithm are used. Also type of output, relation is assumed. Eight registered queries are continuous and registration time of them in system is pre-defined. Queries do not have priority, so number of queries is considered fix. Number of processes is four. Tow parameter are learned by Learning Unit: Time Quantum RR Algorithm ($Q_{RR}$) and Size Buffer ($S_B$).

Time period of execution for the Learning unit is 500 milliseconds. The initial value of parameters which are selected for learning is represented in Table 4. The experiment is done through a 600000 milliseconds period.

Table 4. Initial Value of Parameters

| Parameter | Initialization |
|---|---|
| Time Quantum RR Algorithm | 500ms |
| Size Buffer | 500 tuples |

As we can see in figures 5 and 6, almost after 800 times of execution of LA, both parameters reach a constant value in execution. Table 5 represents achieved values for Quantum length and buffer size after 600000 milliseconds of executing the system.

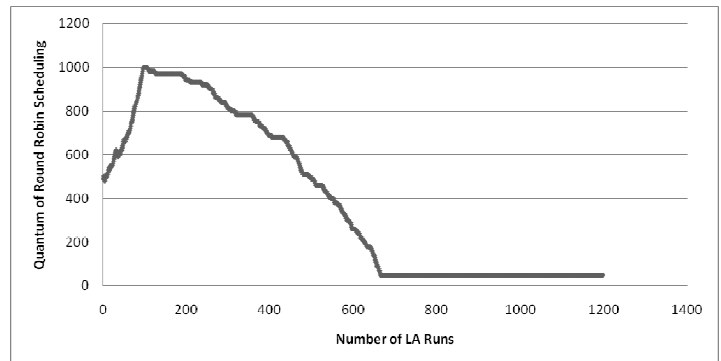

Figure 5 – learning $Q_{RR}$ parameter by LA

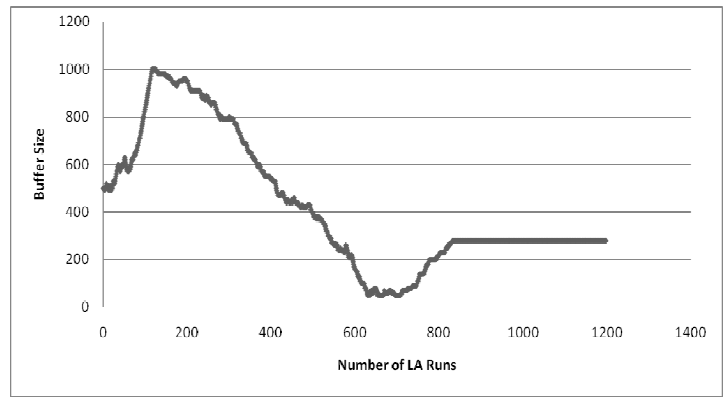

Figure 6. learning $S_B$ parameter by LA

Table 5.  learnt values of parameters

| Parameter | Learned Values |
|---|---|
| Time quantum RR algorithm | 50 ms |
| Size buffer | 280 tuples |

It is expected that the achieved values for parameters which are set by learning to be the optimum values in this execution for the DSMS. To study this as we can see in Table 6, three Metrics of response time, throughput and tuple loss are compared in an execution without learning and another execution using the learning unit. As is seen all three factors are optimized by the learning unit. Response time of the unit is almost reduced to half (Figure 7). The throughput is relatively became more (Figure 8) and the tuple loss (Figure 9) is also decreased.

Table 6. Comparing Performance Metrics

| Performance Metrics | Without Learning | Using Learning |
|---|---|---|
| Response time | 465.521 ms | 220.668 ms |
| Throughput | 1103.546 t/s | 1298.001 t/s |
| Tuple loss | 0.091 | 0.078 |

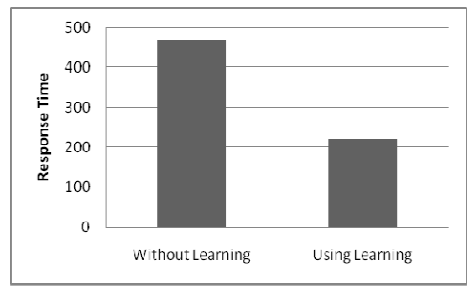

Figure 7. Comparing response time metric in DSMS without Learning and Using Learning.

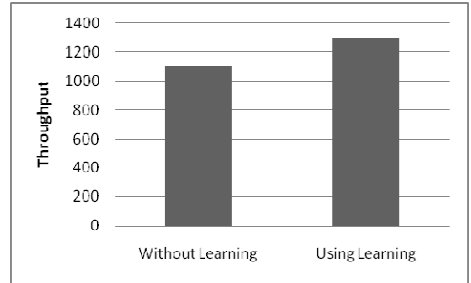

Figure 8. Comparing throughput metric in DSMS without Learning and Using Learning.

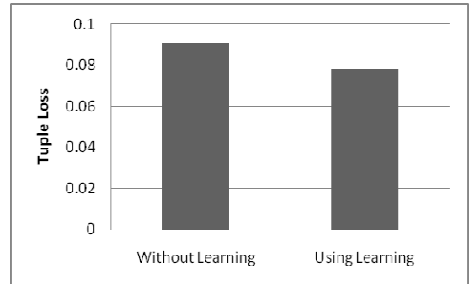

Figure 9. Comparing tuple loss metric in DSMS without Learning and Using Learning.

This comparison shows that when values of parameters of Time Quantum RR Algorithm ($Q_{RR}$) and Size Buffer ($S_B$) are set by the learning unit, considering the features of the stream and the conditions of the system, performance factors of the DSMS will have significant improvements than executions without using the Learning unit.

## 5. RELATED WORKS

Lots of researches on DSMSs are done [13]. Also several primary samples of DSMSs such as the STREAM [1,2], Aurora [5] and TelegraphCQ [14] are provided too.

Some of the available approaches for adaptive query processing in DSMSs are studied. First approach is adaptive query processing without load shedding data. In this approach a generic framework is presented, called as *StreaMon*, for adaptive query processing in a DSMS. The *StreaMon* has three core components: Executor, Profiler and Re-optimizer. In this framework three different composition of continuous type of query which needs adaptivity are represented [2].

In next approach adaptive query processing using the load shedding method is studied. in this approach first applications of load shedding and methods of this approach are studied and then a general framework is provided. Finally two sample architectures of Aurora and Borealis DSMSs after using load shedding for adaptivity of query processing are provided [8].

In next approach the quality adaptive method based on control is studied. In this approach query processing model and quality adaptive framework based on control which includes four elements of Plant, Monitor, Controller and Actuator are provided and then separated study on elements of this framework in addition to defining some functions are provided [26].

scheduling strategies of operators to process continuous queries on data streams varies from simple ones like the Round Robin [6], chain [15] and Greedy [6] to more complex ones [16,17]. Some of them are provided for optimizing a performance factor [18, 19], while some others try to optimize multiple factors or a compound one [20, 21]. Totally most of these methods are provided to minimize tuple latency or the response time factor. Determination and analysis of effective parameters on response time of DSMSs are explicitly studied in few references which [22] is one of them. The learning method is also used as solution to improve adaptivity in systems [24, 25].

## 6. CONCLUSION AND FURTHER WORKS

Performance guarantee in a DSMS is important about static and dynamic features of the system, specifications of input and output data streams, query properties and query processing algorithms. A DSMS can include huge amount of streams and continuous queries. When the system faces some changes of the above, the total performance will decrease. DSMS needs to automatically set of parameters (adptivity) on runtime. The provided daptivity architecture dynamically resets effective and tuneable parameters along the execution using the learning technique, considering the received feed backs from the quality control unit. Defining an internal learning to set time period of the learning unit is a sample of further works. The reason is that in cases which the system is in stable conditions, this unit will be executed in longer periods and while explosive arrival or changes in conditions of the system, it will be executed in shorter periods. Setting the buffer size will be done separately for each of the registered queries.

**Authors**

**Shirin Mohammadi** received her B.S. degree in computer engineering in 2007, She is currently a M.S. student of computer engineering at the Iran University of Science and Technology since 2008. Her research interests include data stream management systems, adaptive query processing, query response time estimation, and real time scheduling.

**Ali A. Safaei** received his B.S. and M.S. degrees in computer engineering in 2001 and 2004, respectively. He is currently a PhD student of computer engineering at the Iran University of Science and Technology since 2005. His research interests include parallel and real-time query processing, quality of services and overload handling in data stream management systems, semantic cache and multiple-query optimization.

**Fatemeh Abdi** received her B.S. and M.S. degrees in computer engineering in 2001 and 2008 at the Iran University of Science and Technology. Her research interests include data mining, mobile databases, stream systems, query processing and Response time estimation.

**Mostafa S. Haghjoo** is an associate professor at the Iran University of Science and Technology. He received his B.Sc. in mathematics and computer science from the Shiraz University in 1976. He received his M.Sc. degree in computer science from the George Washington University in 1978. He obtained his Ph.D. degree in computer science from the Australian National University in 1995.